# RELIABILITY IMPROVEMENT WITH PSP OF WEB-BASED SOFTWARE APPLICATIONS

Leticia Dávila-Nicanor, Pedro Mejía-Alvarez

*CINVESTAV-IPN. Sección de Computación*

ldavila@yahoo.com.mx,
pmejia@cs.cinvestav.mx

***ABSTRACT***

In diverse industrial and academic environments, the quality of the software has been evaluated using different analytic studies. The contribution of the present work is focused on the development of a methodology in order to improve the evaluation and analysis of the reliability of web-based software applications. The Personal Software Process (PSP) was introduced in our methodology for improving the quality of the process and the product. The Evaluation + Improvement (Ei) process is performed in our methodology to evaluate and improve the quality of the software system. We tested our methodology in a web-based software system and used statistical modeling theory for the analysis and evaluation of the reliability. The behavior of the system under ideal conditions was evaluated and compared against the operation of the system executing under real conditions. The results obtained demonstrated the effectiveness and applicability of our methodology.

## 1. INTRODUCTION

Web applications possess different unique characteristics that make web testing and quality assurance different from its corresponding traditional techniques. Web applications can be characterized by the following aspects [5]. Massive Access of users, the simultaneous access of the users in these applications is part of the essence of this type of systems. Web applications provide cross-platform universal access to web resources for the massive user population. For the users it should be transparent that these web applications provide this service to other millions of users. The difficulty of establishing the causes of the errors, since web applications may be access by millions of users, errors have a big impact. Finding the origin of errors in web applications may be difficult and its recovery time may not be immediate, given the great number of software elements that intervene. The integration of diverse software elements for an application in Internet. Web users employ different hardware equipments, network connections, operating systems, middleware and web server support. In a web application, two main components are always required: the backend and the front-end. The backend is the software required for an application in Internet to operate. Among the most important software found in the backend are: the database servers (MySQL, Oracle, Informix, DB2, among those but important), Web Servers (Apache, Netscape Enterprise Server, Netra of Sun, etc.), and the interface programming languages (HTML, XML, PHP, Servlets-Java, Live - Wire, etc.). The front-end is the software required on the part of the client to allow the systems to access the Web Applications. Among the most important software found in the frontend are: Navigators (Explorer, Netscape), which contain plug-in software such as presentations of Macromedia, and languages like JavaScript. Diversity of frameworks to develop, to operate and to maintain a Web site, the development of a Web site requires of a great team of people with different profiles and backgrounds. These teams include programmers, graphic designers, usability engineers, specialists in information integration, network experts and database administrators. This diversity of personnel profiles makes reliable web applications development difficult and sometimes unpredictable. Because of the above characteristics web-based systems tend to evolve rapidly and undergo frequent modifications, due to new technological and commercial opportunities, as well as feedback from the users. In consequence web-based systems are very sensitive to errors. Most work on web applications has been on making them more powerful, but relatively little has been done to ensure its quality. The most important quality attributes demanded by web-based systems are reliability, usability and security. Additional important quality attributes are availability, scalability, maintainability and time-to-market [12]. As with traditional software, verification and validation (V &V) processes have the purpose of checking the quality of the web-based system and revealing non-compliances with the user requirements. However, it is clear that quality assurance should not only include verification and validation but also process improvement. We consider statistical modeling and related reliability analysis in our previous work [18] as one good candidate for effective web quality assurance. This technique help on detecting software errors based on user requirements. In this paper we propose the combination the idea of verification & validation process with Personal Software Process (PSP) as improvement process as part of the quality assurance process of a web-based system. The methodology provides not only a convenient definition on detection software errors, but also to direct the verification and validation process. Structural and thread testing are used to obtain information about the defect density and the mean software fault occurrence in the methodology.

This paper is organized as follows. In Section 2, Web Testing and Quality Assurance is discussed. In Sections 3 a methodology for improving the reliability of web-based software systems is introduced. In Section 4, a case study is introduced to illustrate the use of our methodology. Section 5, we introduce the tool used for the evaluation process. In Section 6, describes the application of the methodology to the case study. Finally in Section 7, we give some concluding remarks.

## 2. WEB TESTING AND QUALITY ASSURANCE

Quality assurance and testing for web applications focus on the prevention and detection of web failures. Web failure is defined as the inability to correctly deliver information or documents required by web users. The following web failure sources are associated with different web layers [11]:

· Host or network failures: Hardware or system failures at the destination host or home host, as well as network failures, may lead to web failures. These failures are mostly related to middleware or web server layers.

· Browser failures: These failures are linked to problems in the web browser at the client side. These failures can be treated the same way as software product failures. Existing techniques for software testing and reliability [1, 10] can be used to assess, predict and improve browser reliability. For web applications there is a strong emphasis in the compatibility among browsers.

· Source or content failures: Web failures can also be caused by the information source itself at the server side. In most cases these kinds of failures reveal non-compliances with user requirements. In our methodology we focus in source failures, instead of browser compatibility or host, network or other browser failures.

According to [8], information source related web components include (a). HTML documents, (b). Java, Javascript, and ActiveX, (c). Cgi-bin Scripts, (d). Database, and (e). Multimedia components. Our goal is to ensure functionality and reliability of this web components and their applications. To do this, we use some forms of functionality testing and thread testing.

## 2.1. Web Modeling and Testing Techniques

Different types of web modeling and testing techniques are being used in the development of web applications. This techniques include, petri nets[9] , model checking[2] and statistical web modeling and testing [5]. Software reliability models the behavior of software systems based on its failures. Predictions such as time to next failure, mean time to failure, or total number of faults detected, are examples of measurements derived from the reliability models. These measurements provide an indicator of reliability growth. The equations for the models have parameters estimated from techniques like least squares or maximum likelihood estimation. Then the equations of the models, often containing exponents or logarithms must be executed. Mathematical and statistical functions provide the predictions and degrees of confidence for the predictions. Verifying that the selected model is valid for the particular data set may require several iterations and an analysis of the model assumptions [17]. According to [5] there are two approaches to statistical web-based software reliability analysis:

· Time domain approach. The failure arrival process is analyzed as a stochastic process using software reliability growth models, which can be used to assess and predict reliability and estimate time or resources needed to reach a reliability target. Markov chains are typically used in this approach. Most web-based applications consists of various components, or stages visible to the users and typically initiated by them. Markov models based on state transitions can generally capture such navigations patterns.

· Input domain approach. This approach is based on repeated error sampling to model software reliability. Input from navigation patterns of web applications is used to test its reliability. We will evaluate the reliability of the system using the input domain approach and study its statistical distribution. The evaluation of the quality attributes selected will be performed using histograms. The main problem is to replace the histograms by theoretical curves or models that represent a probability law. This probability law will allow us to model the behavior of our quality attribute.

# 3. A METHODOLOGY FOR SOFTWARE RELIABILITY IMPROVEMENT.

The main goal in this paper is to introduce a methodology to improve the reliability of web based software systems. This methodology intends to be used to evaluate the quality of web-based software systems using statistical web testing techniques [18]. We will perform an analysis, using statistical modeling techniques, to study the behavior of the system and to detect software defects. Results from our analysis will be evaluated using a case study. We will consider the detection of software defects (errors) resulting from incorrect functionality of the system (i.e., functionality not corresponding to the systems specifications). We intend to use our methodology to evaluate web-based software products already being used on a software industry. The behavior of the system under ideal conditions will be evaluated and compared against the operation of the system executing under real conditions. With the information obtained from this comparison we will be able to acknowledge how far the operation of the system is from the ideal case. After evaluating and predicting the quality of the web based software system, the aim is to introduce the Personal Software Process (PSP) to the software product to improve the quality of the process and the product. After introducing PSP the quality of the product will be again evaluated to measure the reliability and to compare again the operation of the system against the ideal case. The Evaluation + Improvement process will be performed to evaluate and to improve the quality of the software system. The specific phases of the methodology are the following:

1) Evaluation of the quality attribute under ideal conditions. The goal in this phase is to model the system following ideal conditions. In our ideal system the arrivals of the clients into the server are simulated, so that no overload or concurrent problems occur. This evaluation will allow us to obtain reliability measurements for the system executing under ideal conditions.

2) Evaluation of the quality attribute under real conditions. This phase intends to provide quantitative information about the reliability of the real web-based software system. Our case study will be a web-based software for application to graduate courses in a University.

3) Personal Software Process Execution. This phase is introduced as a way to improve the quality of the web-based software system. Since we are applying PSP to well developed web-based software systems, with the application of PSP in our methodology we intend to execute a reengineering process to improve the quality of the software system.

4) Evaluation of the Quality Attribute after PSP. After PSP is implemented, the improvements are measured and compared against the results of the system before PSP. Improvements on the reliability of the system are expected after the introduction of PSP.

5) Analysis of results and conclusions. After applying the Ei process a number of times, we expect to obtain the level of reliability desired. After this process, we propose to evaluate the results and to reach conclusions. Here, we must identify common sources of errors and ways to avoid them. The information obtained from the Ei process will allow also to quantify the effort involved in obtaining desired reliability attributes. This analysis is expected to be useful for future web-based software developments.

## 3.2. Modeling and Evaluation Process

In our methodology, the process required to evaluate the web software product is described in the following steps.

1) Analysis of the initial conditions. The initial conditions of the system are (a). the inputs of the system, (b). its restrictions and non-functional requirements, (c). the services provided by the system and (d). the development conditions. The development conditions allow us to obtain the development process used (i.e cascade, iterative, prototyping, or reuse), the personnel involved in the development, the budget assigned to software development, and the quality standards imposed by the organization.

2) Quality attributes selection and its corresponding metrics. The metrics selected must represent the population under study. In the selection of these metrics is important to choose the appropriate time measurement units. These units describe the time necessary to produce a reliable evaluation process.

3) Measurement Process. A numerical value is the result of this process. This process is responsible for obtaining reliable and easy-to-evaluate results. The steps used in this process are the following.

a. Select the components to be evaluated.
b. Measure the characteristics for the components using the metrics selected.
c. Identify anomalous measurements.
d. Analyze anomalous components.

4) Evaluation and model selection. The procedure required to obtain the statistical model is defined in the following steps.

a. Choose the probability law associated with the population. This law could have an empirical origin. For instance, the histogram that represents the errors occurred in a time interval during the execution of the web system. Examples of these laws are the Gamma distribution, the Poison distribution, the Normal distribution or the Weibull distribution.

b. Evaluate the parameters from the probability law. A probability law contains parameters that depend on the population under study. Modeling is used as a way to provide the mathematical equations necessary to obtain the values of these parameters. Different techniques from numerical analysis can be used to estimate the parameters of the model. For instance, maximum likelihood estimates (MLE's), least squares method, or polynomial regression [7]. It may be necessary to consider the use of several techniques to obtain the parameters of the model. Once the parameters are obtained, the distribution function graph from the model must be drawn to observe its behavior and its tendency. This graph will denote the behavior of the quality attribute.

c. Compare the probability law. Once chosen the probability law, and its corresponding parameters we must verify that the law chosen is in accordance with the population under study. If the result from this comparison is favorable, then we can be certain that the law chosen represents the population under study. Otherwise, we may have to choose another probability law and continue with the procedure from the beginning. In this step, test results for reliability assessment and prediction help with decision making. The results from this evaluation are graphs known as histograms, where the x-axis denotes the values of the metric used and the y-axis denotes its frequency. This histograms will allows us to approximate the probability model that best represent the behavior of the population (metrics) under study. After this histogram is developed, it must be replaced by the probability law that best represent its behavior. This law will represent the behavioral pattern of the metrics under study and will indicate the correctness of our evaluation.

5. Model Validation. With the parameters obtained we must corroborate that the model represent the population of metrics under study. In this step the histogram is compared against the distribution function obtained.

6. Overall evaluation. In this step we must evaluate the information provided by the evaluation process (information relevant to the quality attribute). Predictions about the future behavior of the quality attribute can also be assessed in this step.

# 4. CASE STUDY

In this section we will introduce a case study to illustrate our methodology. Our case study will be a Web-based Software for an On-line application to Graduate courses in a University (SOGU). The architecture of the system is based on a Linux platform (Red Hat V.8)[13] using an Apache web server [15] and a MySQL database management system[14]. The interface language used is PHP[16]. The number of lines of code used was 1200. The development time was 6 months. The block diagram of the SOGU system is illustrated in Fig. 1. In our case study there are three main views: (a). Professors, (b). Students and (c) General Public Users. Access to the system is granted by the use of a password. Each type of user is allowed to view only a dedicated part of the system. In general this system is capable of allowing students to register for courses in a given University, and also allow Professors to register their courses, verify the data provided by the students and provide grades for their courses. Students and Professors can Insert, Delete, Update and Read information from their own databases. General public users can only read courses information. This system is composed of 3 different databases. The courses database, the students database and the Professors database. The system controls the access of different users and manages the information flow provided to the users. Changes to databases are allowed only during specific time periods (e.g., start of the terms).

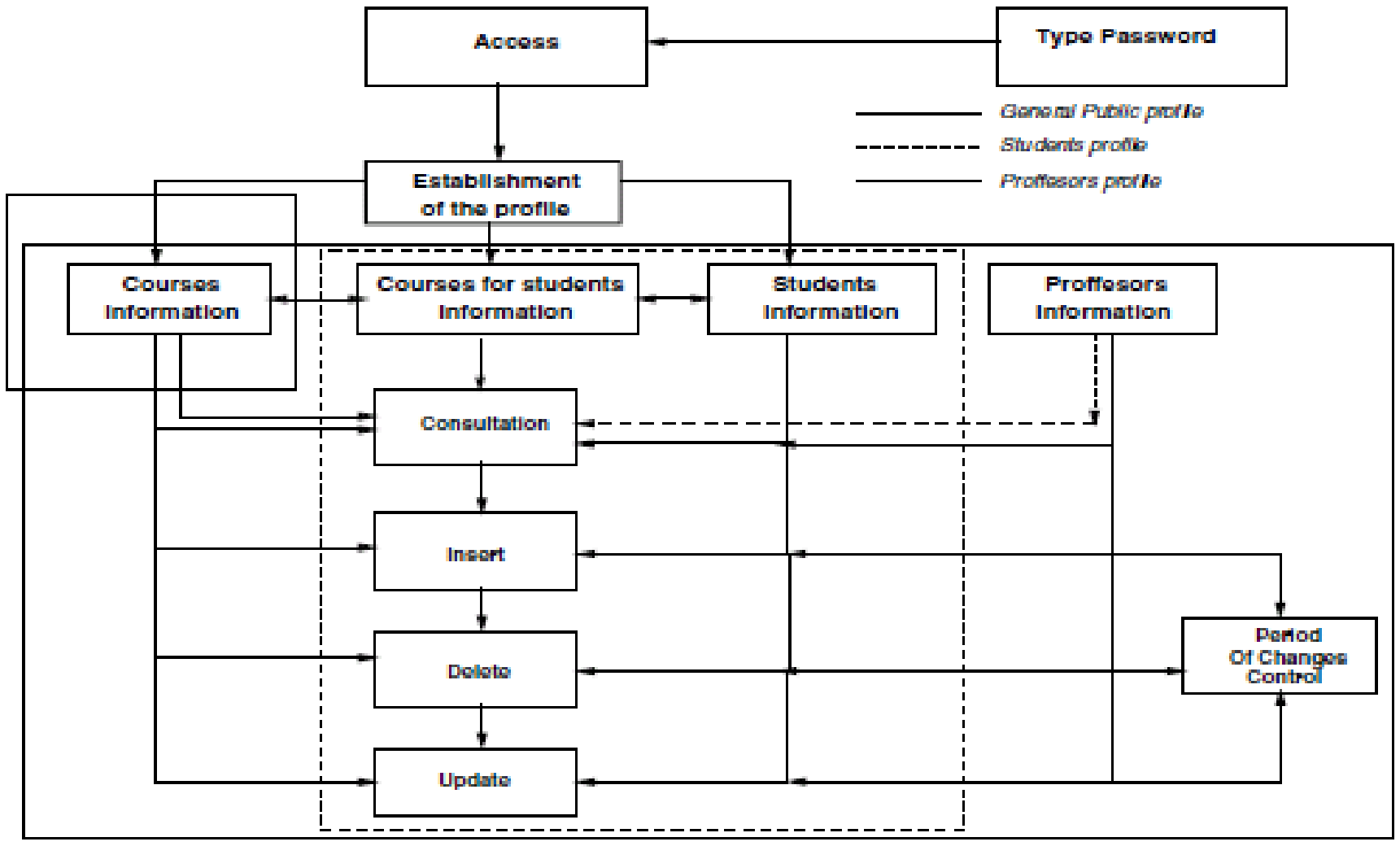

Fig. 1 Modules from the SOGU System

# 5. A TOOL FOR THE EVALUATION PROCESS

Testing a web-based software system is different from testing traditional software systems. In a traditional software system there are no multiple accesses to the system, and single testing software can test the system. In a web-based software there may be frequent accesses from many users using different computing platforms, therefore web-based testing must ensure that concurrent testers validate the execution of the system. We developed a Testing Tool for our SOGU system in Java (under a Linux platform). This tool is used to evaluate the quality attribute under real conditions. Evaluation is conducted by our testing tool using concurrent test threads (system testers). Each test thread is responsible for executing a specific functionality test specified on its corresponding test case file. Test cases are generated randomly and test data is prepared to perform functionality tests. On each test case an specific test profile indicates an specific path of navigation (type of test) and the view that the tester will test. The test thread, have access to test cases (which contain the test data and the test profile) and the activity log files. The activity log files are files that contain the activities performed by the test thread. The analyzer reads the activity log files and produces an error log file which contains the specific faults detected and the defect density computed. In our tool, it is possible to perform multiple simulations and compute averages from the results of the simulations.

As shown in Fig. 2, our testing tool was built in three modules, the initialization module, the tester module and the analyzer module. The initialization and the tester modules were developed in Java, while the analyzer module was developed using scripts. The initialization module, build the structure model of the web site to evaluate. That is, it finds all navigation paths of the web site. Also, this module is responsible for activating the system testers. Any error found during the testing process is recorded into an HTML error-log file. The second module, perform the navigation and testing of the SOGU system, according to its web-site structure and its test case file. The third module analyzes the data obtained from the testing process and computes the defect density and the mean time to failure (MTTF).

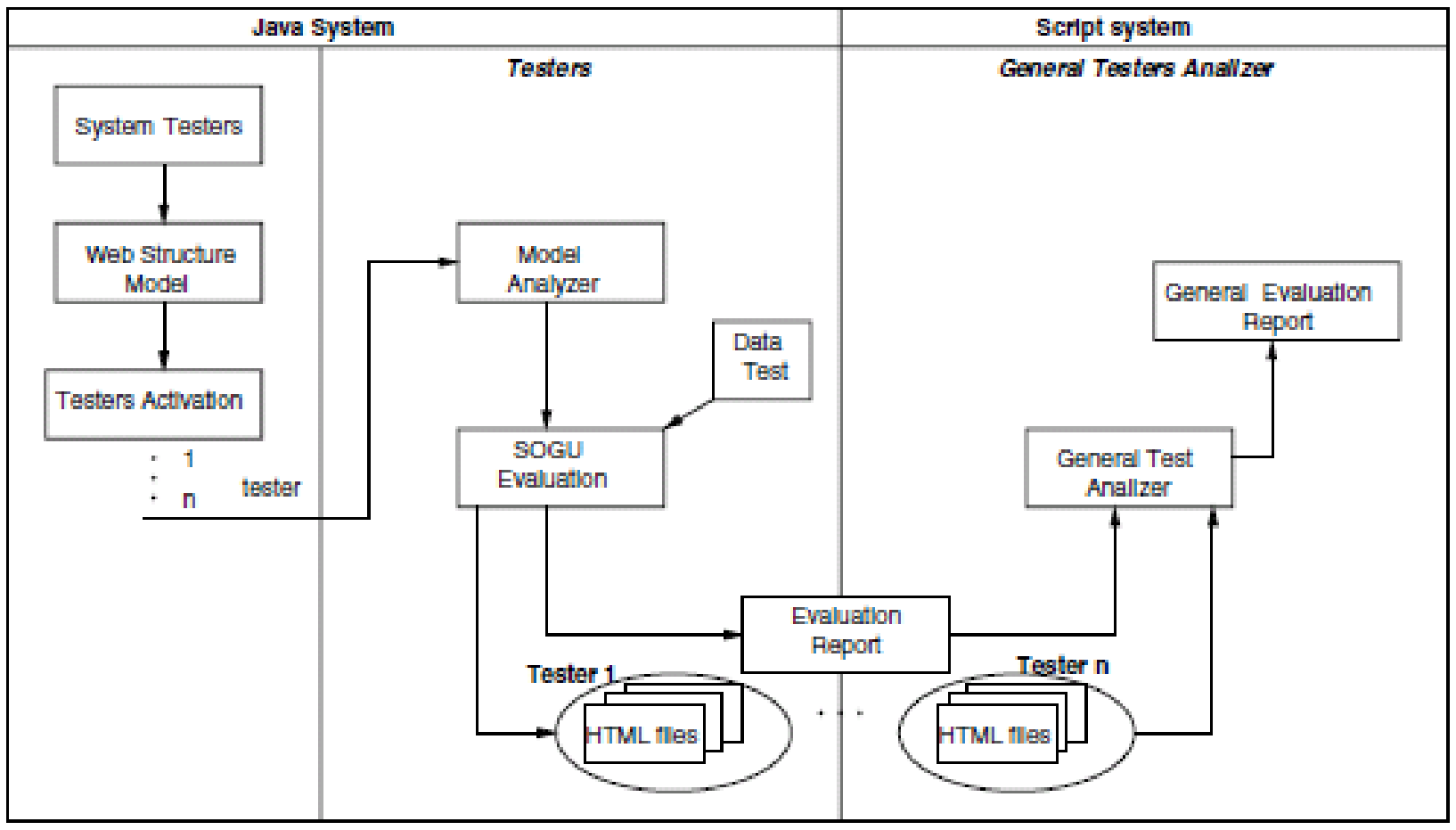

Fig. 2 Modules of the Evaluation Tool

# 6. EVALUATION OF THE QUALITY ATTRIBUTE

## 6.1. Evaluation of the quality attribute under ideal conditions

1) Initial Conditions under the ideal case. A simulator was developed in Java to test the quality attribute under ideal conditions. The simulator uses a producer-consumer model (shown in Fig. 2, where the information is handled using a queue of shared resources. Synchronization between producer and consumer is performed using thread synchronization. The producer generates the input data for the consumer. The functionality of the consumer consists on updating and drawing the behavior of the simulation at each time unit, based on the data provided by the producers shared queue. The producer is composed of 5 modules. In the main module, SOGU process, the execution cycle of the simulator is executed. The second module, Initialization Routine, initializes the structures, the statistic counters and the simulation of the first arrival. The third module, arrive, calculates the user arrival’s times, verifies if the users are allocated in the web server queue and verifies if the system has enough capacity for such users. If the system has enough capacity, then the access is granted to a given user and a call is performed to the Addeparture function. In the fourth module, AddDeparture, the departure function is called for detecting the errors generated during the simulated execution. In case some errors are detected, the defect density is increased allowing the user to exit the system. The fifth module, departure, computes the finishing exit times for each user for its corresponding thread of execution. Since the ideal case is simulated, no overload or concurrency problems occur. The system is tested under these ideal condition over our SOGU system. The modeling of the user arrivals is performed using M/M/1 queues on the web server with a probability of $(n + 1)-1$, where n denotes the size of the queue (i.e., number of user in the system). The exit condition for any user is (a) when an error is detected during the operation of the system, or (b) when the user requests its departure from the system. Since our web-server is based on a Intel 500 MHz processor with 70 GBytes of disk and 500 Mbytes of RAM memory, the capacity of the web system is set to 100 concurrent users. The histogram illustrated in Fig. 4 is obtained after computing the average values of 500 simulations. For each simulation, 100 discrete events are simulated. For each event a pre-determined navigation path is specified. As discussed before, there are 3 types of users, students, professors and general public. An event is executed by a user that is allowed to execute for t time units. The value of t is generated following a normal distribution with mean μ = 3 seconds, and variance σ = 1 seconds. The time between user arrivals is generated following an exponential distribution with mean μ = 4 seconds.

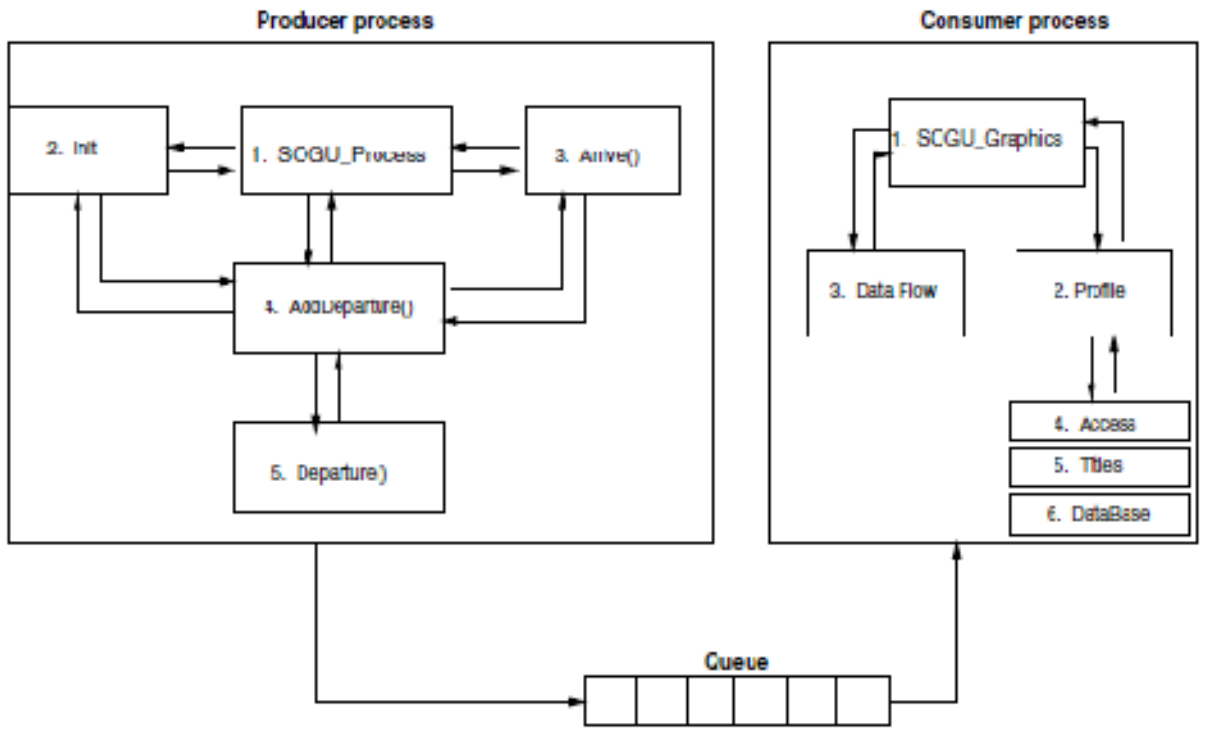

Fig. 3 Producer and Consumer Diagram

2) Quality attribute selection and its corresponding metrics.
We are interested in evaluating the reliability attribute. Reliability is a quality attribute that evaluates the degree of fault-free operation of the system. We are interested in evaluating the defect density as our reliability metric. The defect density denotes the number of errors detected in a given time interval.

3) Measurement Process. In the simulations all components were tested. The metric used was the defect density.
During the execution of the simulation only 5 from 505 measurements were discarded for being anomalous.

4) Evaluation and model selection. The histogram resulting from the evaluation process is shown in Fig. 4. Note that the histogram has a tendency to approximate a Weibull distribution curve.

a. Choose the probability law associated with the population. The probability function used to represent our histogram is the following:

$$f(x) = \alpha\beta^{-\alpha}x^{\alpha-1}e^{-(x/\beta)^{\alpha}} \qquad (1)$$

The probability function chosen is the Weibull distribution. The Weibull distribution has been used in various engineering fields for reliability analysis[6].

b Evaluate the parameters from the probability law. The values of α and β were obtained in [7] using maximum likelihood estimators by the following Equations:

$$\frac{\sum_{i=1}^{n} X_i^{\alpha} lnX_i}{\sum_{i=1}^{n} X_i^{\alpha}} - \frac{1}{\alpha} = \frac{\sum_{i=1}^{n} lnX_i}{n} \qquad (2)$$

$$\beta = \left(\frac{\sum_{i=1}^{n} X_i^{\alpha}}{n}\right)^{1/\alpha} \qquad (3)$$

where n = 500 (number of simulations) and Xi denotes the defect density values obtained from each simulation and illustrated in the histogram of Fig. 4. Equation 2 is solved using the Newton-Raphson numerical method while Equation 3 is solved directly using the α value. The values obtained from the above Equations are: α = 1.63 and β = 2.4. Substituting the α and β values in Equation 1 we obtain,

$$f_i(x) - \begin{cases} 0.391x^{0.63}e^{-(x/2.4)^{1.63}} & \text{If } x \geq 0 \\ 0 & \text{Otherwise} \end{cases}$$

Based on Equation $f_i(x)$, we illustrate the performance of the reliability attribute on Fig. 5.

5) Validation of the model. From Fig. 5, it can be noted that the Weibull distribution chosen is capable of modeling the reliability attribute, thus validating the process. We can say that the defect density is an adequate metric capable of representing the reliability of our web-based software product.

6) Overall validation. From Fig 5, it can be noted that low values of defect density are found under a high frequency. It is possible to note that defect density maximum values are between 0 and 2, while the maximum frequency values are less than f(x) = 0.35. The defect density values obtained are between 0 and 8. In this simulations we conclude that the model used was capable of modeling the reliability of the ideal web-based system.

## 6.2. Evaluation of the quality attribute under real conditions.

1) Initial conditions under the real case. The Initial conditions for the real case are as follows. The evaluation of the real case is performed using the testing tool described in previous section. The views of the execution testers were: Professors, Students and General public. All modules from the SOGU system were evaluated using functional testing. A test case file was generated randomly, containing a large number (1000) of threads of execution (specific navigation paths) associated with a given execution profile. The web server used consisted of a PC executing at 500 MHz. In the real case, 500 evaluations were conducted. The average of the results from the simulations for the real case are shown in the Histogram illustrated in Fig. 6. For each evaluation the testing time was 100 seconds. The time between the arrival of each tester was computed following a exponential distribution with mean equal to μ = 4 seconds. Each tester is allowed to execute for t time units. The value of t is generated following a normal distribution with mean μ = 3 seconds, and variance σ = 1 seconds.

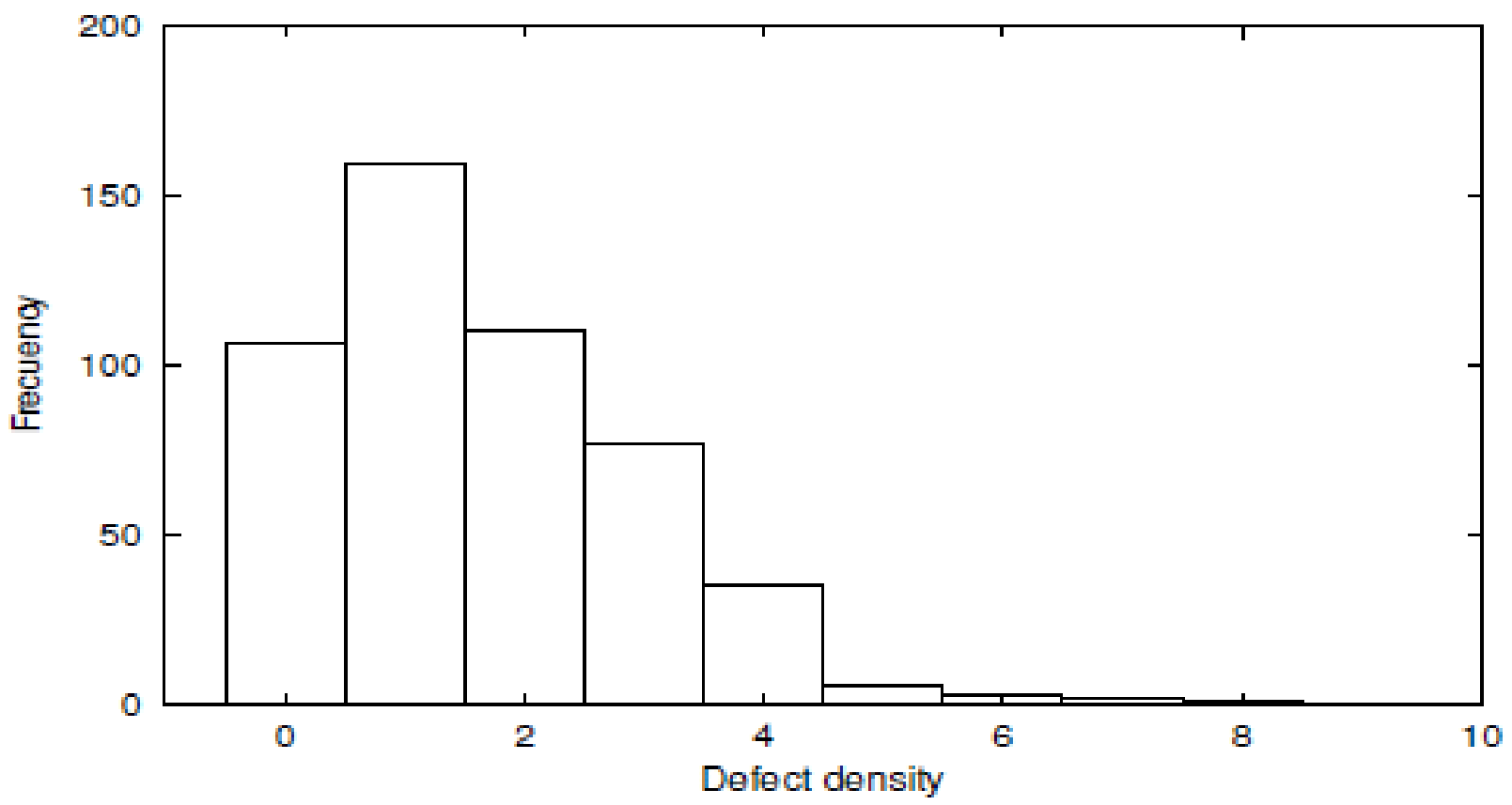

Fig. 4 Histogram for the Ideal Case

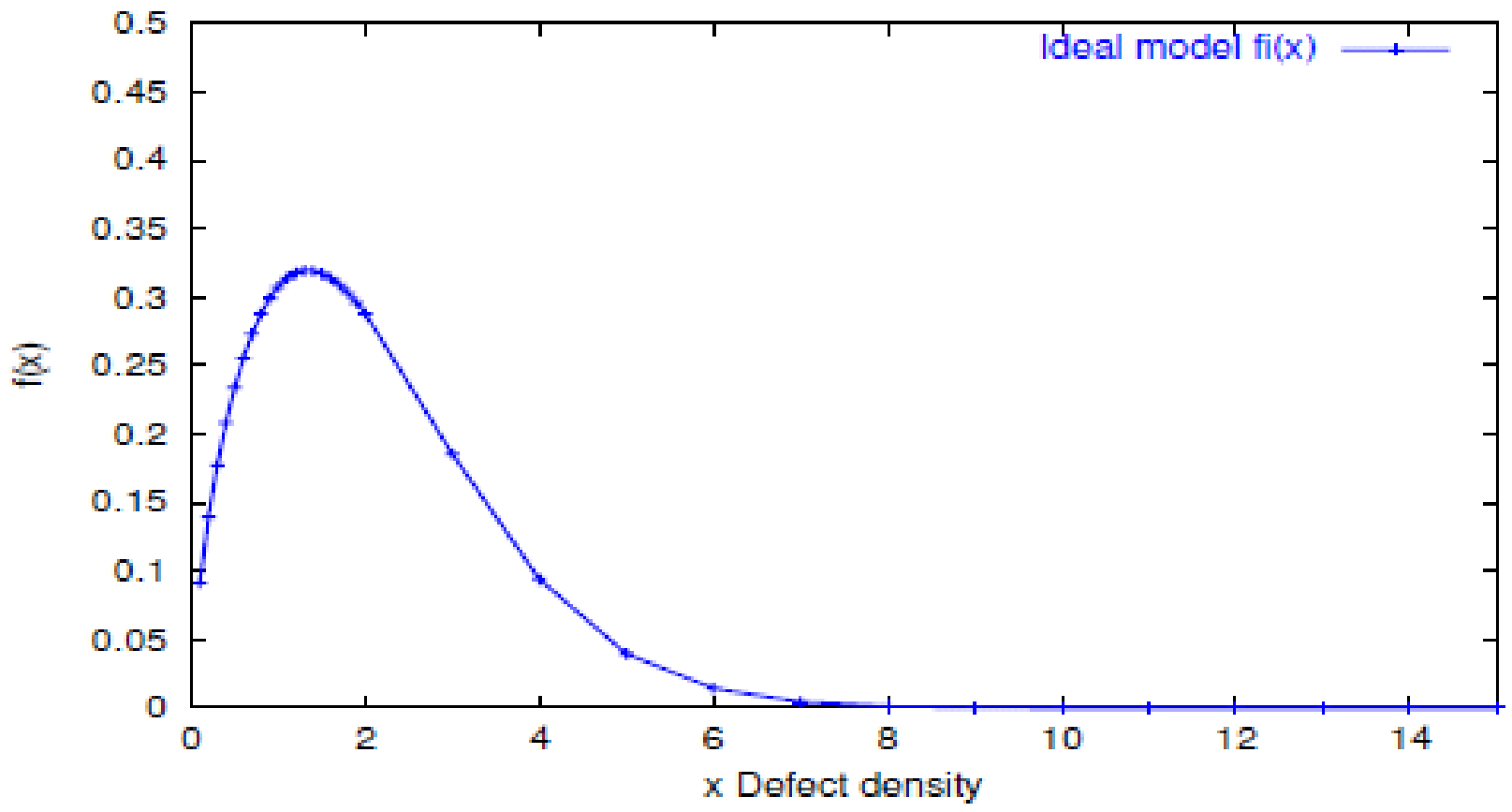

Fig 5 Defect Density for the Ideal Case

1. Quality attribute selection and its corresponding metrics. As in the ideal case the quality attribute to evaluate is the reliability, and its metric was the defect density.

2. Measurement Process. All modules from the SOGU system were tested. During the execution of the system only 3 evaluations were discarded for being anomalous. The specific errors found on each of the modules from the SOGU system are not detailed in the paper because of space restrictions.

3. Evaluation and model selection. The results shown in the histogram of Fig. 6 indicate for example show that for 63 evaluations a defect density of errors of 10 (10 errors occurred in average). From the Histogram we can note that the highest frequency of errors occur with a density of 10, and the smallest frequency of errors occur with a density of 1, 26 and 32. The probability law associated with the population and its corresponding α and β parameters are the same as those described in Equations 1, 2 and 3 respectively. The values obtained from the Equations are: α = 2.16 and β = 12.8. Substituting the α and β values in Equation 1 we obtain,

$$f_i(x) = \begin{cases} 0.00876x^{1.16}e^{-(x/12.8)^{2.16}} & \text{If } x \geq 0 \\ 0 & \text{Otherwise} \end{cases}$$

Based on Equation fi(x), we illustrate the performance of the reliability attribute for the ideal and real cases in Fig. 7. It is possible to note from the figure that the results obtained from the real case improve the results obtained from the ideal case.

### 6.3. PSP Implementation

The Personal Software Process [3] is a process improvement methodology aimed at individual software engineers. It claims to improve software quality (in particular defect content), effort estimation capability, and process adaptation and improvement capabilities. We have tested some of these claims in our SOGU web-based system to compare the defect reduction before and after PSP. Our aim is to include PSP in our Ei process discussed before.

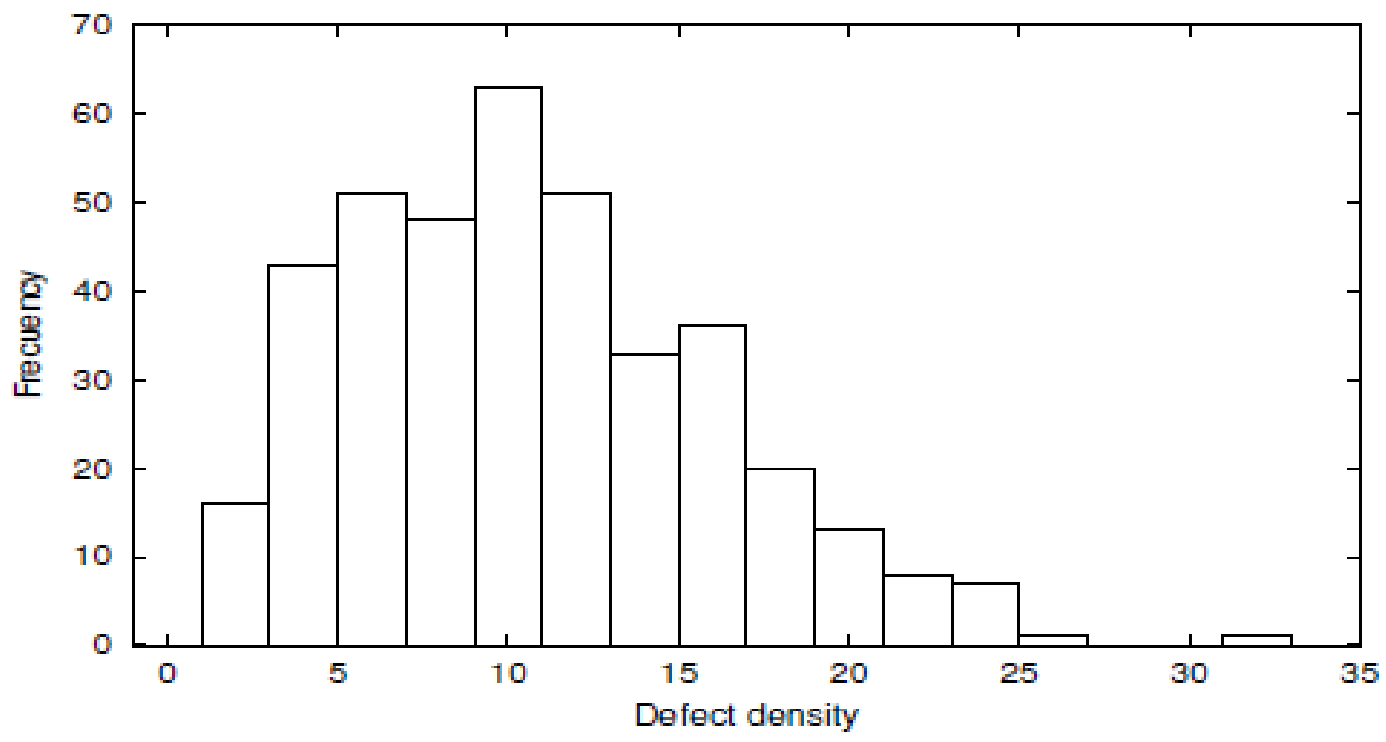

Fig. 6 Defect Density Histogram for the Real Case

The goals of the PSP are that individual Software Engineers learn:

- How to accurately estimate, plan, track and re-plan the time required for individual software development.
- How to work according to a well-defined process, and how to re-define this process.
- How to use reviews effectively and efficiently for improving software quality and productivity.
- How to avoid software defects.

- How to analyze measurement data for improving estimation, defect removal and defect prevention.

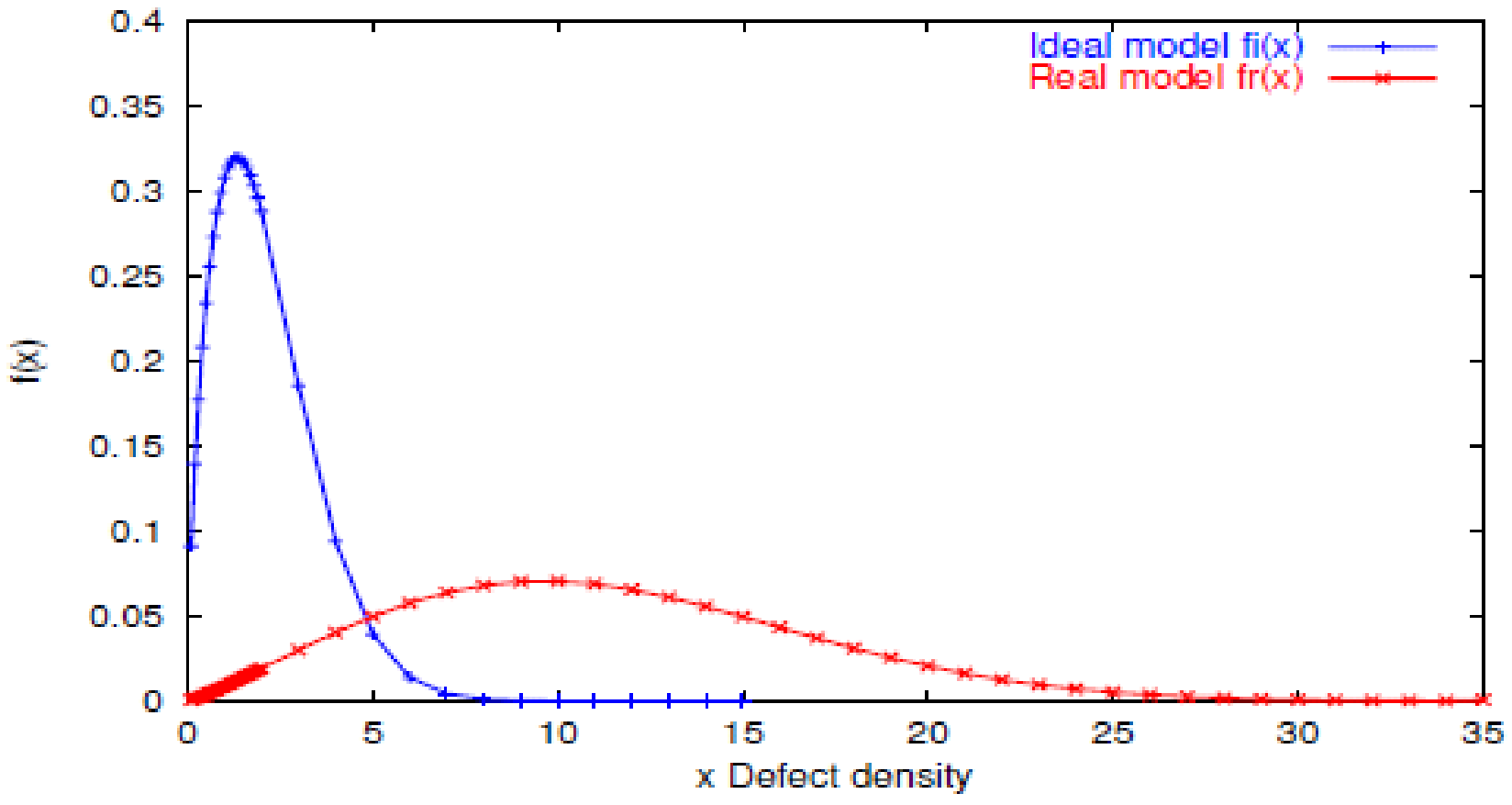

Fig. 7 Graph for the Ideal and Real Case

PSP consists of a series of scripts that define tasks or forms for recording data, and standards that govern such things as coding practices, size counting, and the assignment of defect types. The first step in PSP is to first plan the work ahead and document the plan. As this work is done, the development times are recorded and every defect found is tracked and reported. At the end of the project, PSP includes a postmortem analysis and a complete project plan summary report. PSPs quality improvements result from three key aspects: First, by tracking all defects, software developers are sensitized to the mistakes they personally make and therefore become more careful in their work. Second, when they analyze their defect data, they gain a clearer understanding of the cost of removing defects and thus apply the most effective ways of finding and fixing them. And third, PSP introduces a number of quality practices that have proved effective in preventing defects and in efficiently finding and fixing them. PSP has been implemented after the phase of evaluation of the quality attribute under real conditions in our SOGU system. The improving in reliability will be measured by comparing the evaluation procedure before and after PSP implementation. In the application of PSP into our SOGU web-based system only 1 Engineer (without previous experience in applying PSP) was involved. The results obtained from the implementation of PSP [4] are illustrated in Fig. 8,9,10, 11 and 12. Since we were only interested on detecting defects, other results obtained from PSP, are not shown in this paper.

In Fig. 8 the yield versus program number is illustrated. Yield is the principal PSP quality measure. Total process yield is the percentage of defects found and fixed in the web-based software system. From this figure we conclude that the yield tend to improve while our Engineer applied PSP to more programs. Defect trends are shown in Fig. 9. While the defect trend fluctuated while developing more programs, we can observe that in general there is was a reduction of defects/KLOC while PSP was applied to more programs. In Fig 10 and Fig. 11, we illustrate the defects elimination rate and the defects introduction rate respectively. From these figures, it is possible to observe a sharp decrease in the defects elimination and introduction rate while applying PSP to more programs. Finally, Fig 12 shows the improvement in A/FR from the application of PSP to our SOGU system. From the above results we conclude that the application of PSP allowed us to detect and eliminate many defects from our SOGU system, and that most of the defects detected (and eliminated) were found on the coding phases.

## 6.4. Evaluation of the Quality Attribute after PSP

After implementing PSP for the SOGU system several improvements were detected. This section is devoted to the presentation of the results obtained for the evaluation of the quality attribute after PSP.

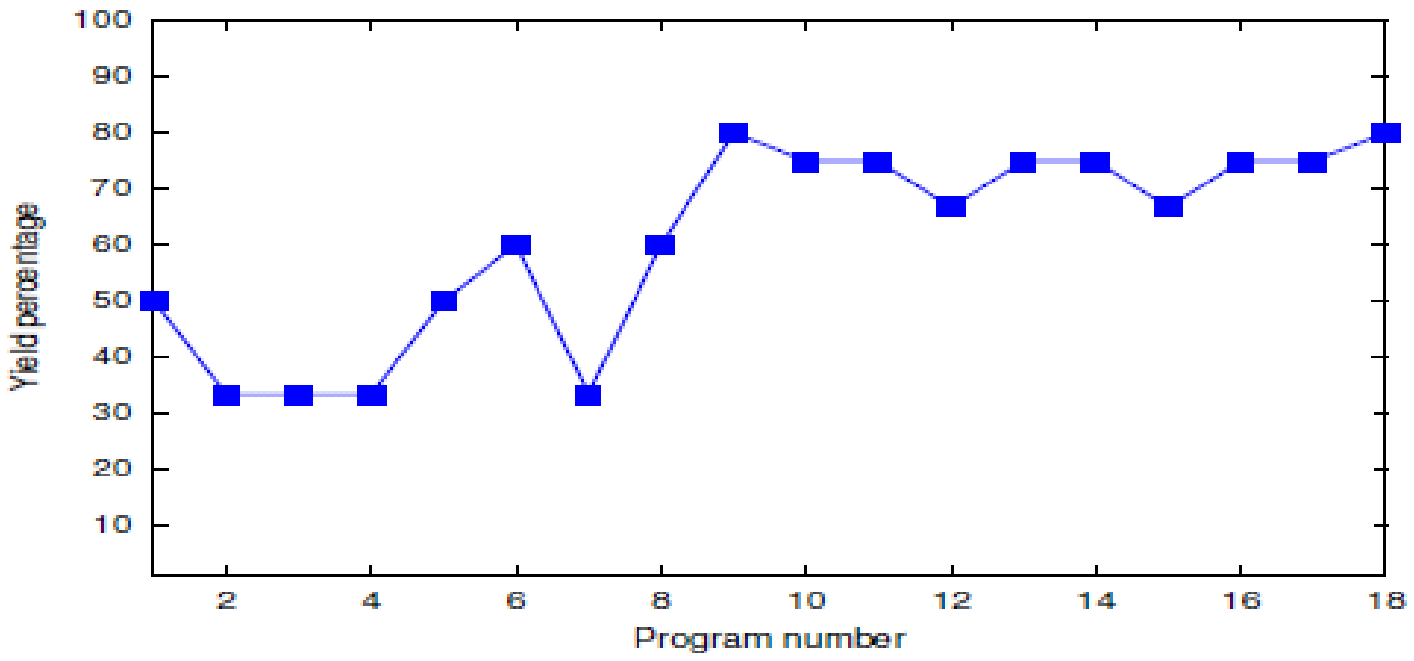

Fig 8 Yield versus program number

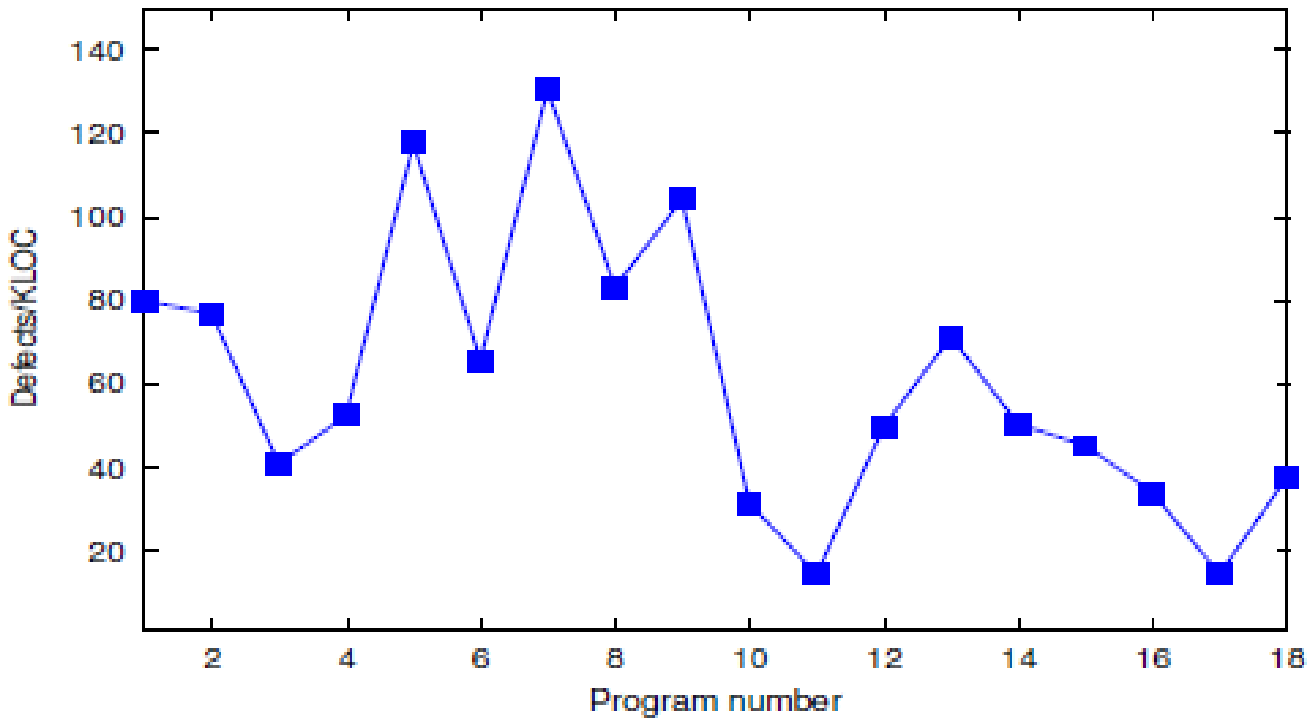

Fig 9 Defects per KLOC trend

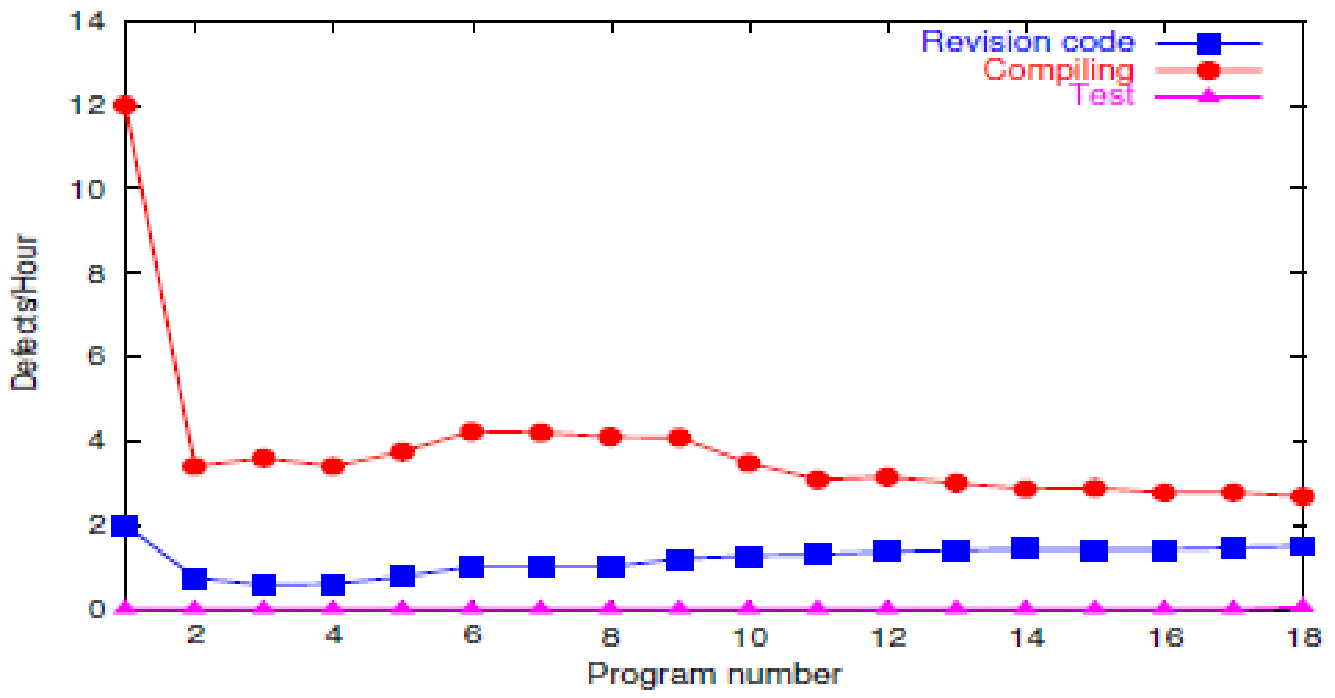

Fig 10 Defects elimination rate

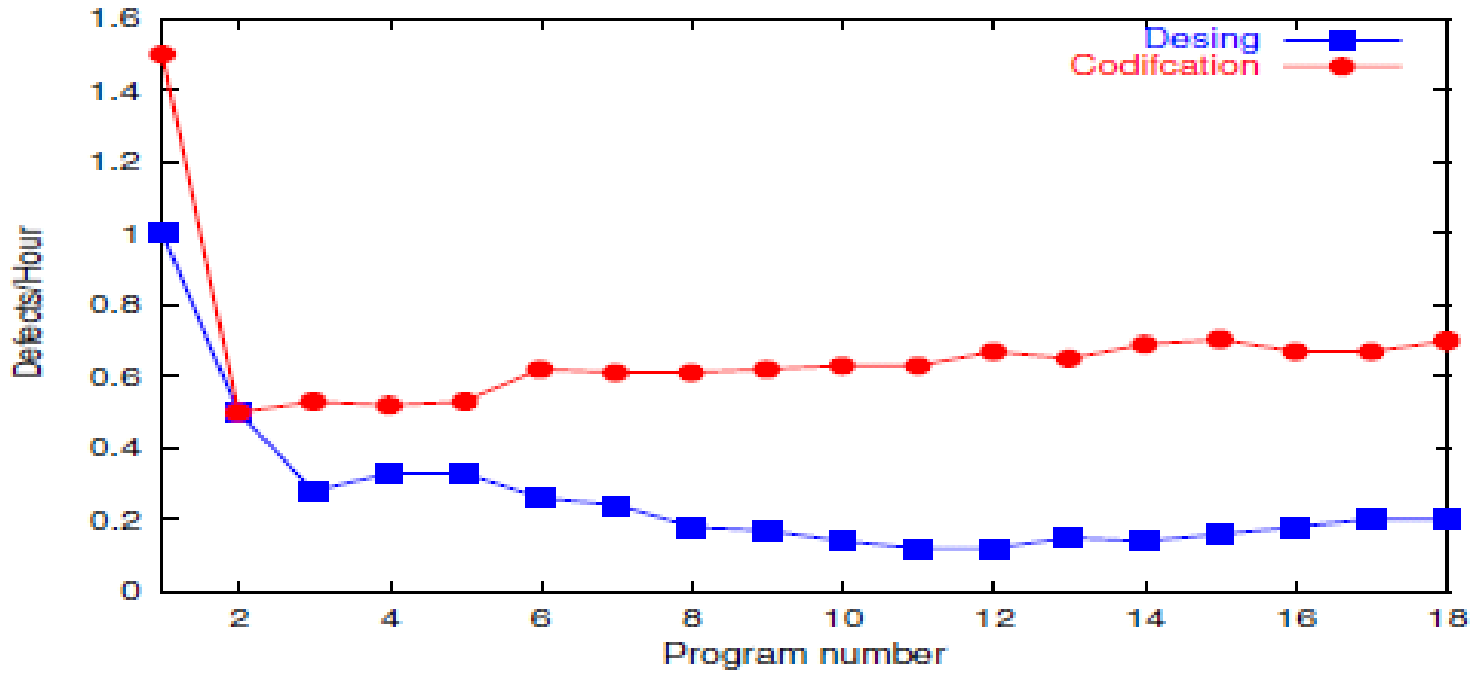

Fig 11: Defects introduction rate

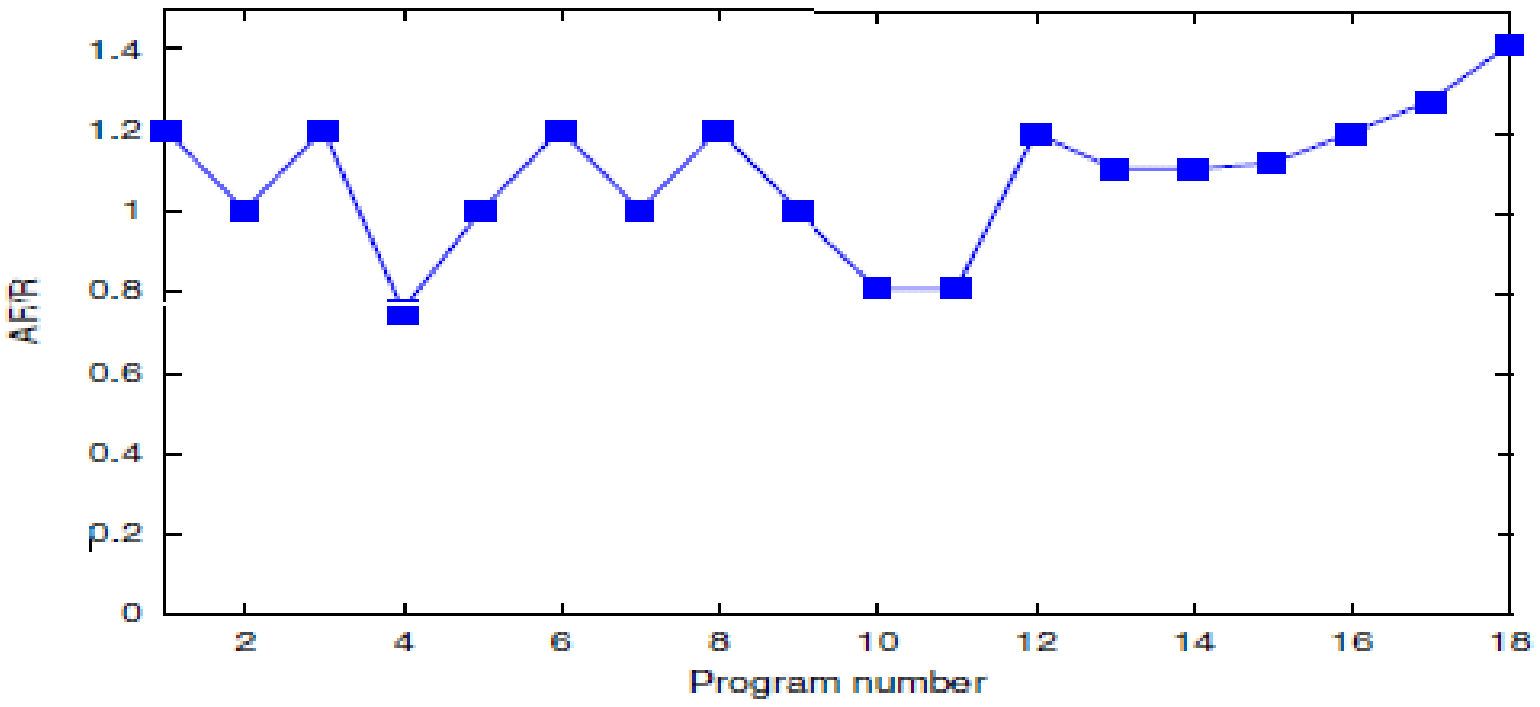

Fig 12. AF/R versus program number

Phases 1, 2 and 3 from the evaluation process were similar to those from the evaluation before PSP.

4. Evaluation and model selection.

Fig. 13 illustrates the Histogram produced after implementing PSP. The probability law associated with the population and its corresponding α and β parameters are the same as those described in Equations 1, 2 and 3 respectively. The values obtained from the Equations are: $\alpha = 1.26$ and $\beta = 5.19$.

Substituting the α and β values in Equation 1 we obtain, Proceedings of

$$f_i(x) = \begin{cases} 0.158x^{0.26}e^{-(x/5.19)^{1.26}} & \text{If } x \geq 0 \\ 0 & \text{Otherwise} \end{cases}$$

Based on Equation fi(x), we illustrate the performance of the reliability attribute on Fig. 14. From Fig. 14 it is possible to observe that the reliability of the SOGU system after implementing the PSP process improves. The curve for the model after PSP indicates that now the reliability is near the ideal case. The conclusion from this evaluation process is that the reduced defect density observed in figure 14 for the model indicates that fewer errors were found in the system after implementing PSP.

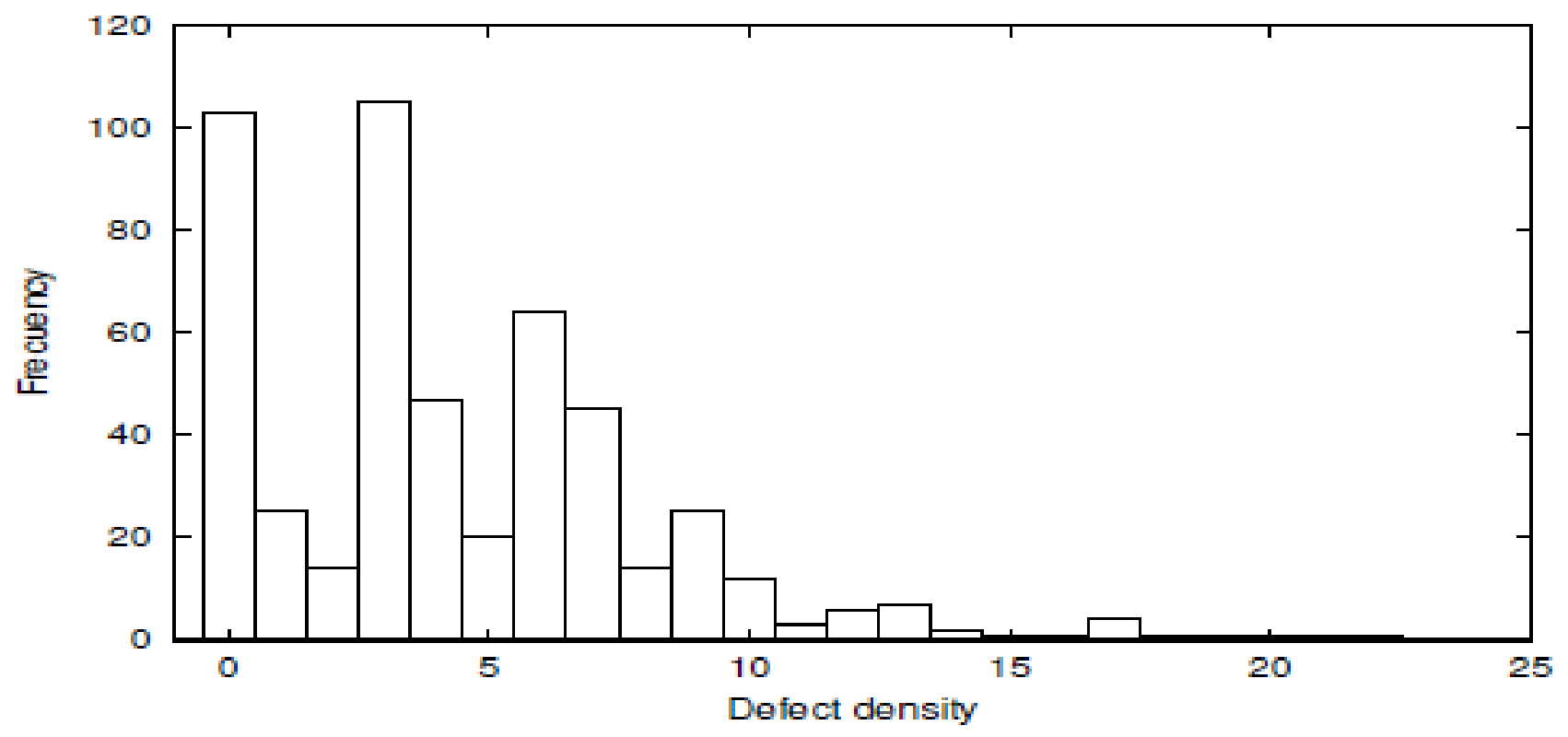

Fig. 13 Defect Density Histogram for the Real Case after PSP

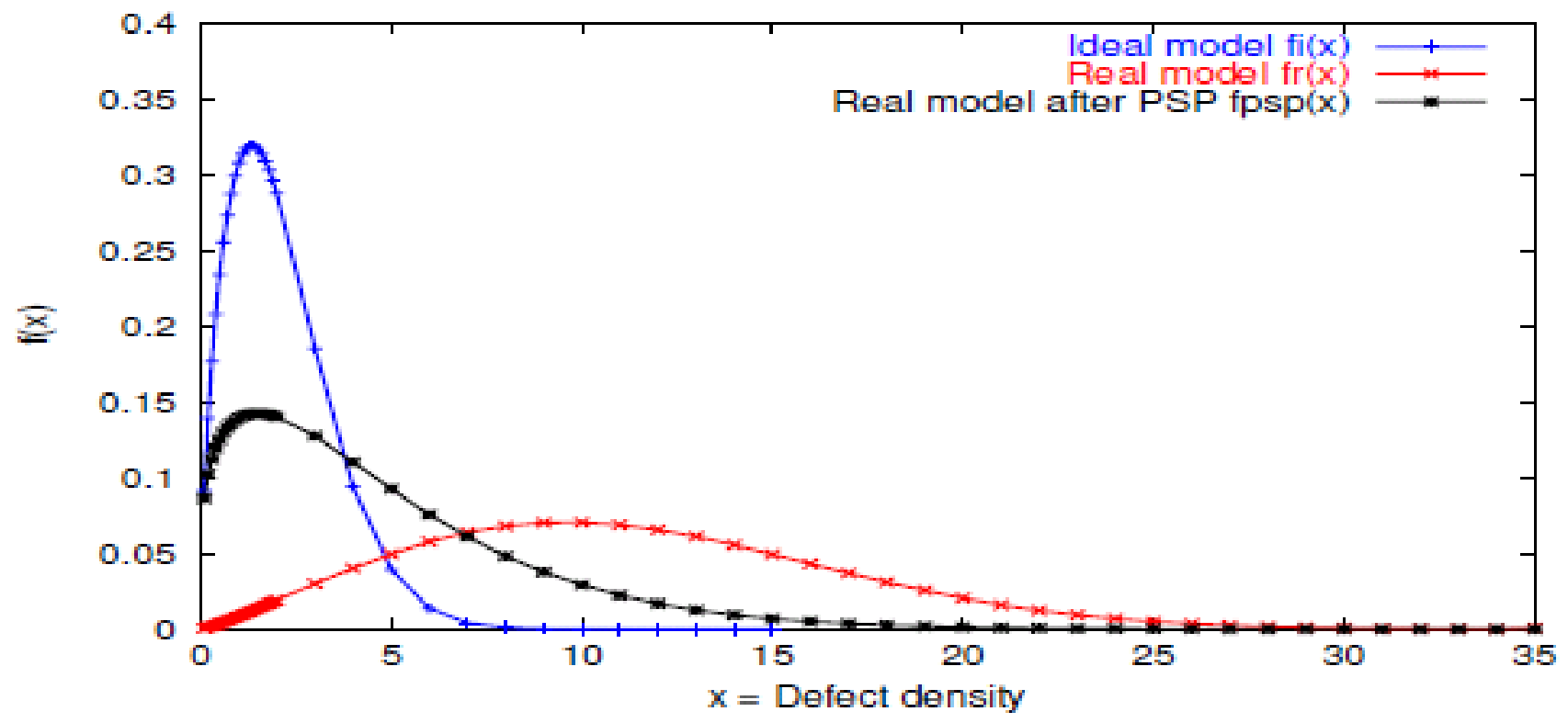

Fig. 14 Graph from the Real Case after PSP

# 7. CONCLUSIONS

In this paper we introduce a methodology for reliability testing and improvement of web-based software applications. This methodology, introduce a V &V + Process improvement process as part of the quality assurance process of a web-based system. The Personal Software Process (PSP) is used in our methodology to improve the reliability of the web-based software application. We apply statistical modeling and related reliability analysis [6] for effective web quality assurance. These techniques help on detecting and fixing software errors based on user requirements. Structural testing is used to obtain information about the defect density and the mean software fault occurrence. A Testing Tool was developed for our SOGU system to evaluate the quality attribute under ideal and real conditions. We tested our methodology on Web-based Software for an On-line application to graduate courses in a University (SOGU). Overall, we conclude that this methodology provides an effective framework for assessing and improving reliability of web-based software applications. Our future goals are to test our methodology in a more complex web-based software application, and to extend it to include the markov-chain statistical testing models. Since our testing tool was developed ad-hoc for our web-based application we plan to develop a more general testing tool that integrate our methodology.